\documentstyle[prl,aps,twocolumn]{revtex}
\begin{document}
\draft
%\preprint{to be submitted to ............ }
\title{Subdiffusive and superdiffusive quantum transport and
generalized duality}
\author{Maura Sassetti$^1$, Henning Schomerus$^2$, and Ulrich
Weiss$^3$}
\address{
${}^1$Istituto di Fisica di Ingegneria, INFM,
Universit\`{a} di Genova, I-16146 Genova, Italy\\
${}^2$ Fachbereich Physik, Universit\"at--Gesamthochschule Essen,
 D-45117 Essen, Germany\\
${}^3$Institut f\"ur Theoretische Physik, Universit\"at Stuttgart,
 D-70550 Stuttgart, Germany }
\date{Date: \today}
\maketitle
\begin{abstract}
As a generic model for transport of interacting fermions through a
barrier or interstitials in a lattice, quantum Brownian motion in a
periodic potential is studied. There is a duality transformation
between the continuous coordinate or phase representation and the
discrete momentum or charge representation for general
frequency-dependent damping. Sub-Ohmic friction is mapped on super-Ohmic
friction, and vice versa. The mapping is exact for arbitrary barrier
height and valid at any temperature. Thus all features of the
continuous model can be investigated from analytical or
numerical analysis of the discrete model. New nonperturbative results
for the frequency-dependent linear mobility including subdiffusive and
superdiffusive behaviors are reported.
\end{abstract}


\narrowtext
Quantum Brownian motion in a periodic potential (QBM) is a generic
model for many transport phenomena in condensed matter \cite{fi}.
Major themes in the past have been the phenomena of quantum
coherence and localization. Transport of charge in one-dimensional
interacting Fermi systems through barriers or impurities
within the continuous $\theta$-phase representation of
the Luttinger liquid model \cite{kane} is also described by the above
model. The harmonic liquid in the fermionic model away from the barrier
corresponds to the thermal reservoir in the QBM model, and the conductance
in the first model corresponds to the mobility in the latter. Short-range
electron-electron interaction with coupling parameter $g$
($g<1$ in the repulsive case) corresponds to Ohmic damping in the
connected QBM model, and $g$ is related to the Ohmic damping parameter
$\alpha$ by $g = 1/\alpha$. Moreover, unscreened long-range Coulomb
repulsion \cite{schulz} corresponds to sub-Ohmic damping in the related
QBM model \cite{fabr}, leading to strong suppression of the conductance
at low temperatures. A variety of other physical or chemical
systems involves transport of charge through barriers for Ohmic or
super-Ohmic damping \cite{leggett87,kag92}.

Schmid has shown for Ohmic damping \cite{schm} that the continuous
QBM transport problem at $T=0$ can be mapped on a
tight-binding model. The mapping resulted in a duality
transformation for the dc mobility in which diffusive and localized
behavior are interchanged. The mapping was generalized
later to finite $T$ \cite{fi}. Similarly,
the continuous phase representation of the fermionic problem is
directly related to a discrete charge representation.

In this Letter, we generalize the mapping to frequency-dependent
damping, e.g., unscreened Coulomb interactions in the fermionic problem
\cite{schulz,fabr}. Our study will reveal an exact mapping between
the respective Hamiltonians. Then, upon using the equations of motion,
an exact duality relation between the respective frequency-dependent
linear mobilities or conductances will be derived.
New results exhibiting super-diffusive, diffusive, sub-diffusive, and
strictly localized behaviors, depending on the parameter regime, will be
given. Here we restrict the attention to the QBM transport problem, as
the transcription into the fermionic problem is straightforward
\cite{kane}.

As another essential reason for our study, it should be stressed
that it is important to perform numerical
real-time simulations of the dynamics within the equivalent discrete model.
The discrete variables significantly reduce the relevant configuration
space subject to Monte Carlo sampling compared to the continuous
model, and meaningful real-time simulations in the interesting
nonperturbative low-temperature regime are possible \cite{leung}.

The system-plus-bath Hamiltonian for a Brownian particle of mass $M$
moving in a washboard potential (WB) with period $Q_0$ and
corrugation strength $V_0$ is
\begin{eqnarray}\label{hwc}
H_{\rm WB}^{} &=& P^2/2M -
V_0\cos\left(2\pi Q/ Q_0 \right) \\
&& + \sum_{\alpha}\left[\frac{p_{\alpha}^2}{2m_{\alpha}}+
\frac{m_{\alpha}\omega^2_{\alpha}}{2}\left(x_{\alpha} -
\frac{c_{\alpha}}
{m_{\alpha}\omega^2_{\alpha}}Q\right)^2\right]\;. \nonumber
\end{eqnarray}
Here we have included the usual potential renormalization term
$ \sum_\alpha (c_\alpha^2/2m_\alpha\omega_\alpha^2)Q^2$
which provides invariance of the  coupling
$\sum_\alpha c_\alpha x_\alpha Q$ under spatial translations.
As far as we are interested in properties of the particle, the coupling
constants $c_\alpha$ and the parameters of the bath are relevant only
via the spectral density of the coupling
$ J_{\rm WB}^{}(\omega)= (\pi/2)\sum_{\alpha}
 (c^2_{\alpha}/m_{\alpha}\omega_{\alpha})
\delta(\omega-\omega_{\alpha})\,.$

The dual tight-binding (TB) model with transfer
matrix element $V_0$  and translational-invariant coupling is
\begin{eqnarray}\label{htb}
H_{\rm TB}^{} &=& - \frac{V_0}{2} \left( e^{i q_0 P /\hbar}\, +\,
{\rm h.c.} \right) \\
&& + \sum_{\alpha}\left[\frac{\pi_{\alpha}^2}{2M_{\alpha}} +
\frac{M_{\alpha}\Omega^2_{\alpha}}{2}\left(u_{\alpha}-
\frac{d_{\alpha}}
{M_{\alpha}\Omega^2_{\alpha}} Q \right)^2\right]\; , \nonumber
\end{eqnarray}
\[
Q = q_0 \sum_n n a_n^\dagger a_n^{}\; ;\quad
  e^{iq_0 P/\hbar} = \sum_n a_n^\dagger a_{n+1}^{} \; ,
\]
where $q_0$ is the TB lattice constant. In the discrete representation,
the operators $a_n^\dagger$ and $a_n^{}$ create and annihilate a particle at
the site $n$, respectively.
The spectral density of the environmental coupling in the
tight-binding model is
$ J_{\rm TB}^{}(\omega)= (\pi/2) \sum_{\alpha} (d^2_\alpha
 /M_{\alpha} \Omega_{\alpha})\delta(\omega-\Omega_{\alpha})\, .$

In the first step of the mapping of (\ref{hwc}) onto (\ref{htb}),
we perform the canonical transformations
($ \kappa = 2\pi\hbar/Q_0 q_0 $)
\begin{eqnarray*}
 p_\alpha &\to& -m_\alpha \omega_\alpha x_\alpha \, ,
\quad  x_\alpha \to  p_\alpha/m_\alpha\omega_\alpha +
 P c_\alpha/\kappa m_\alpha\omega_\alpha^2  \, ,\\
Q &\to& P/\kappa \, ,\quad \qquad\quad\,  P \to -\kappa Q  +
\sum_\alpha c_\alpha x_\alpha/\omega_\alpha \, .
\end{eqnarray*}
In the intermediate form of the Hamiltonian, the modes of the environment
are coupled with each other. The corresponding term of the Hamiltonian is
\begin{eqnarray*}
% H_{\rm WB}^\prime &=& \frac{\kappa^2 Q^2}{2M}  -
% V_0\cos\left(\frac{q_0 P}{\hbar}\right)
% -\frac{\kappa Q}{M}  \sum_\alpha
% \frac{c_\alpha x_\alpha}{\omega_\alpha} + H_{\rm R}^{}\, ,\\
H_{\rm R}^{} &=& \sum_\alpha \left[\frac{p_\alpha^2}{2m_\alpha}
+ \frac{m_\alpha \omega^2_\alpha x^2_\alpha}{2}\right]
+\frac{1}{2M}\left(\sum_\alpha \frac{c_\alpha
x_\alpha}{\omega_\alpha} \right)^2  .
\end{eqnarray*}
In the next step, we diagonalize $H_{\rm R}^{}$, thereby introducing new
canonical variables $\pi_\alpha$ and $u_\alpha$, and new parameters
$M_\alpha$ and $\Omega_\alpha$. By the transformation, the interaction term
$(\kappa Q/M)\sum_\alpha c_\alpha x_\alpha/\omega_\alpha$ is changed
into $Q\sum_\alpha d_\alpha u_\alpha$.

The mapping of (\ref{hwc}) onto (\ref{htb}) is completed on imposing
spatial invariance of the coupling $  \sum_\alpha d_\alpha u_\alpha Q $,
\begin{equation}\label{iden2}
\kappa^2/M = \sum_\alpha d^2_\alpha / M_\alpha \Omega^2_\alpha \; .
% }=\frac{2}{\pi}\int_0^\infty \!
% {\rm d}\omega \frac{J_{\rm TB}^{}(\omega)}{\omega}\; ,
\end{equation}
By (\ref{iden2}), the mapping is constrained, as we shall see.

The relation between the spectral functions $J_{\rm WB}^{}(\omega)$ and
$J_{\rm TB}^{}(\omega)$ results from properties of the dynamical matrix
$A_{\alpha\beta}(\omega)$ of the reservoir Hamiltonian
$H_{\rm R}^{}$. We have $ A_{\alpha\beta}(\omega) = B_{\alpha\beta}(\omega) +
 c_\alpha c_\beta /M\omega_\alpha\omega_\beta\, $,
where $B_{\alpha\beta}(\omega)=
\delta_{\alpha\beta}m_\alpha(\omega_\alpha^2 -\omega^2)$. Observing that the
coupling terms are in the form of an exterior vector product, it is
straightforward to calculate the ratio of the determinants of
the matrices $A$ and $B$,
$\Delta^{-1}(\omega)\equiv \det A(\omega)/\det B(\omega)$,
\begin{eqnarray} \nonumber
% \frac{\det A(\omega)}{\det B(\omega)} \equiv
\Delta^{-1}(\omega)
  &=& 1 + \frac{1}{M}\sum_\alpha
 \frac{c^2_\alpha}{m_\alpha\omega^2_\alpha(\omega_\alpha^2 -  \omega^2)}  \\
 &=&  1 + i\gamma_{\rm WB}^{}(\omega)/\omega  \; . \label{dadb}
\end{eqnarray}
In the second line, we have introduced the spectral damping function
$\gamma_{\rm WB}^{}(\omega)$. The function $\gamma(\omega)$ is defined in
terms of the continuous spectral density of the coupling $J(\omega)$ of the
respective model by
\begin{equation}\label{kwc}
 \gamma(\omega) = \lim_{\epsilon\to 0^+}
\frac{2\omega}{i\pi M}
\int_0^\infty \! d\omega'
\frac{J(\omega')}{\omega'(\omega'^2-\omega^2-
i\epsilon\,{\rm sgn}\,\omega)} \; .
\end{equation}
On the other hand, we may resolve $A_{\alpha\beta}$ for $B_{\alpha\beta}$
and then switch to the unitarily equivalent form in which $A$ is diagonal,
$\widetilde{A}_{\alpha\beta}(\omega) = \delta_{\alpha\beta}
M_\alpha (\Omega_\alpha^2 -\omega^2)$. We then have
$ \widetilde{B}_{\alpha\beta}(\omega) = \widetilde{A}_{\alpha\beta}
(\omega) - (M/\kappa^2) d_\alpha d_\beta \,$.
{}From this we find
\begin{eqnarray}\nonumber
% \frac{\det B(\omega)}{\det A(\omega)}
 \Delta(\omega) &=& 1 - \frac{M}{\kappa^2}\sum_\alpha
 \frac{d_\alpha^2}{M_\alpha(\Omega_\alpha^2-\omega^2)} \\
 &=&  - i (M^2/\kappa^2)\, \omega \gamma_{\rm TB}^{}(\omega) \; ,
\label{dbda}
\end{eqnarray}
where in the second line we have used (\ref{iden2}), and where
$\gamma_{\rm TB}^{}(\omega)$ is the damping function of the TB model.
Combining (\ref{dadb}) with (\ref{dbda}), we obtain an exact relation
between the damping functions of the two models,
\begin{equation}\label{gammarel}
\gamma_{\rm TB}^{}(\omega)
\left[ \gamma_{\rm WB}^{}(\omega)- i\omega \right] =
\kappa^2/M^2   \; .
\end{equation}
With the relation $J(\omega) = M\omega\,{\rm Re}\,\gamma(\omega)$,
we then have
\begin{eqnarray}\label{jjwc}
 J_{\rm WB}(\omega) &=& (\kappa^2/M^2) J_{\rm
TB}(\omega)/ |\gamma_{\rm TB}^{}(\omega)|^2 \; ,\\ \label{jjtb}
J_{\rm TB}(\omega) &=& (\kappa^2/M^2) J_{\rm
WB}(\omega)/ |\omega + i\gamma_{\rm WB}^{}(\omega)|^2\; .
\end{eqnarray}
Thus, by use of (\ref{kwc}), the spectral density of the one model
can be calculated for any form of the spectral density of the other model.

To be definite, let us assume a simple power-law form for the spectral
density of the WB model in the entire relevant frequency range,
\begin{equation}\label{lowfr}
 J_{\rm WB}(\omega) = \eta_s \widetilde{\omega}
 (\omega/\widetilde{\omega})^s \; , \qquad 0<s<2 \; ,
\end{equation}
resulting in the spectral damping function
\begin{equation}\label{gammaess}
 \gamma_{\rm WB}^{}(\omega) =
\gamma_s (-i \omega/\widetilde{\omega})^{s-1}
\end{equation}
with $\gamma_s = \eta_s/[M\sin(\pi s/2)]$.
Here we have introduced for $s\neq 1$ a reference frequency
$\widetilde{\omega}$ so that the coupling strength
$\gamma_s$  has the usual dimension of frequency.
The case $s=1$ describes (frequency-independent) Ohmic dissipation,
while the cases $s<1$ and $s>1$ are usually referred to as sub-Ohmic
and super-Ohmic friction.

We are now in the position to make a check on the consistency of the
condition (\ref{iden2}). There follows from (\ref{dbda}) that the
constraint (\ref{iden2}) is satisfied if $\Delta(\omega)$ vanishes at
zero frequency. Now, upon inserting (\ref{gammaess}) in
(\ref{dadb}), we see that the condition $\lim_{\omega\to 0}\Delta(\omega)
\to 0$ is satisfied for $s$ in the range $0<s<2$.
For $s\ge 2$, we have $\gamma_{\rm WB}^{}(\omega\to 0)\propto \omega$.
Damping becomes ineffective in this limit, and the only
remaining effect is mass renormalization \cite{gra}.

We have thus demonstrated  an exact duality between the weak
corrugation model (\ref{hwc}) and the tight-binding model
(\ref{htb}). In the mapping, the continuous coordinate $Q$ in the WB model is
identified, up to a scale factor $1/\kappa$, with the quasimomentum $P$
in the dual TB model. The mapping is possible only for nonvanishing
dissipation since otherwise the scale factor is infinity. Strictly speaking,
the mapping holds under the condition
$\lim_{\omega\to 0} \omega^2/J(\omega) \to 0$.
There are no restrictions on
the form of $J(\omega)$ at finite frequencies except those
enforced on physical grounds. There follows from (\ref{jjwc}) or
(\ref{jjtb}) that the spectral density
$J_{\rm WB}^{}(\omega) \propto \omega^s$ of the weak corrugation
model maps on the spectral density
$J_{\rm TB}(\omega\to 0)\propto \omega^{s^\prime}$ of the dual TB
model, where $s^\prime =2 -s$.
Thus formally in the spectral densities of the coupling, the power
$s$ is mapped on the power $2-s$.
% Correspondingly, the spectral
% damping function $\gamma_{\rm WB}^{}(\omega) \propto \omega^{s-1}$
% of the WB model maps at low frequencies onto the power-law behavior
% $\gamma_{\rm TB}^{}(\omega)\propto \omega^{1-s}$ of the dual model.
Physically, this means that sub-Ohmic and super-Ohmic friction are
interchanged in the transformation, while Ohmic friction is mapped
on Ohmic friction.

Upon inserting (\ref{lowfr}) into (\ref{kwc}) and using (\ref{jjtb}),
the spectral density of the dual TB model takes the form
\[
J_{\rm TB}(\omega) = \frac{\kappa^2}{\eta_s}
\frac{\widetilde{\omega}
(\omega/\widetilde{\omega})^{2-s}}{ 1 +\left[\cot(\pi s/2) -
(M\widetilde{\omega}/\eta_s)(\omega/\widetilde{\omega})^{2-s}
\right]^2} \; .
\]
In contrast to the chosen form for $J_{\rm WB}(\omega)$,
the function $J_{\rm TB}(\omega)$ exhibits a soft cutoff.
Physically, this is due to the finite mass of the particle in the
washboard potential model which provides a natural cutoff in the spectral
density. In the Ohmic case, the expression for
$J_{\rm TB}^{}(\omega)$ reduces to the Drude form \cite{fi},
$ J_{\rm TB}(\omega) \propto \omega/[1 + (\omega/\gamma_1^{})^2]\,$.

After having completed the mapping of the Hamiltonians, we turn to a
study of the respective linear ac mobilities. First,
we observe that by the above unitary transformations a coordinate
autocorrelation function of the WB model is transformed into a
momentum autocorrelation function of the associated TB model. From
this we find for the linear mobility of the WB model
\begin{equation}\label{mobwc}
 \mu_{\rm WB}^{}(\omega) = -i\omega X_{\rm TB}^{}(\omega)/\kappa^2 \; ,
\end{equation}
where $X_{\rm TB}^{}(\omega)$ is the Fourier transform of the
retarded momentum response function of the TB model.
On the other hand, the ac mobility of the TB model is
related to the Fourier transform $Y_{\rm TB}^{}(\omega)$ of the respective
retarded coordinate response function of the TB model by
\begin{equation}\label{mobtb}
 \mu_{\rm TB}^{}(\omega) = -i\omega Y_{\rm TB}^{}(\omega) \; ,
\end{equation}
To relate $X_{\rm TB}^{}(\omega)$ to $Y_{\rm TB}^{}(\omega)$, we use
the equations of motion resulting from (\ref{htb}).
% Upon eliminating $u_\alpha(t)$, we find
% \[
% \dot{P} (t) = \int_{-\infty}^\infty\! \! {\rm d}\tau\,
% K (t-\tau) Q(\tau) + \sum_\alpha d_\alpha
% u_\alpha^{(0)}(t) \; ,
% \]
% where $u_\alpha^{(0)}(t)$ describes free bath motion, and where
% $K (t)$ is the retarded integral kernel with Fourier transform
% $K(\omega)= iM\omega\gamma_{\rm TB}^{}(\omega)$
% with $\gamma_{\rm TB}^{}(\omega)$ given by (\ref{kwc})
% for $J(\omega)=J_{\rm TB}^{}(\omega)$.
% We then find that $X_{\rm TB}^{}(\omega)$ is related to
% $Y_{\rm TB}^{}(\omega)$ by
This yields
\[
 \omega^2 X_{\rm TB}^{}(\omega) = iM\omega\gamma_{\rm TB}^{}(\omega)
 - M^2 \omega^2 \gamma_{\rm TB}^2(\omega) Y_{\rm TB}^{}(\omega) \; .
\]
Using this, and the relations (\ref{mobwc}), (\ref{mobtb})
and (\ref{gammarel}), we find for the
linear ac mobility the exact relations
\begin{eqnarray}\label{dualwb}
 \mu_{\rm WB}^{}(\omega) &=& \frac{M}{\kappa^2} \gamma_{\rm
TB}^{}(\omega)
-\frac{M^2}{\kappa^2}\gamma_{\rm TB}^2(\omega)\mu_{\rm
TB}^{}(\omega)\;, \\ \label{dualtb}
\mu_{\rm TB}^{}(\omega) &=& (M/\kappa^2)[\gamma_{\rm WB}^{}
(\omega) -i\omega] \\
&& -  (M^2/\kappa^2) [\gamma_{\rm WB}^{}(\omega)
-i\omega]^2 \mu_{\rm WB}^{}(\omega) \;. \nonumber
\end{eqnarray}
The expressions (\ref{gammarel}) -- (\ref{jjtb}), (\ref{dualwb}) and
(\ref{dualtb}) are the central relations of the duality.
Since they have been derived by using canonical transformations and
commutation relations,
they are not restricted to the discrete representation in
(\ref{htb}).
Most remarkably, the duality holds for {\em any spectral form} of the damping
kernel, with the only restriction that damping remains effective at
zero frequency. Thus, there is a far more general duality between these
models than considered in earlier studies \cite{fi,schm,schon}.
Here, the correspondence is shown for the
frequency-dependent linear mobility. It is important that no
restrictions were placed on $T$ and on $V_0$.

So far, we have considered the length scales $Q_0$ and $q_0$ as free
parameters. In order to simplify the relation between the
ac mobilities, we now make the specific choice
\begin{equation}\label{qqrel}
 M\gamma_s Q_0 q_0 / 2\pi\hbar \equiv M\gamma_s/\kappa = 1 \; .
\end{equation}
Then the standard dimensionless coupling parameters
$\alpha_{\rm WB}^{} =: M\gamma_s Q_0^2/2\pi\hbar$ and
$\alpha_{\rm TB}^{} =: M\gamma_s q_0^2/2\pi\hbar$ are related by
$ Q_0/q_0 = \alpha_{\rm WB}^{} =1/ \alpha_{\rm TB}^{} \, .$

In the remainder of the discussion, we confine ourselves to the
low-frequency regime $\omega/\gamma_{\rm WB}^{}(\omega) \ll 1$, in which
the relations (\ref{dualwb}) and (\ref{dualtb}) become symmetric.
Upon using (\ref{gammarel}), (\ref{gammaess}) and
(\ref{qqrel}), we obtain the relation $(\mu_0^{} =: 1/M \gamma_s)$
\begin{equation}
 \mu_{\rm WB}^{}(\omega) =
(i\widetilde{\omega}/\omega)^{s-1} \mu_0^{}
  - (i\widetilde{\omega}/\omega)^{2s-2}
\mu_{\rm TB}^{}(\omega)\; . \label{duality}
\end{equation}

For $s=1$, this reduces to the duality relation of the Ohmic case,
$\mu_{\rm TB}^{}(\omega) +\mu_{\rm WB}^{}(\omega) = \mu_0^{}$.
Most analytic work on Ohmic friction has been restricted to the TB model with
a very high cutoff frequency of the bath modes. Much of the theoretical
interest has been concentrated on the localization transition at $T=0$ at
friction strength $\alpha_{\rm TB}^{} =1$ and on the renormalization group
flows. For a discussion of this, and of the
implications of the duality relation for the WB model, we refer to the
literature \cite{fi,kane,gui85,zwe}. In many interesting cases, e.~g. charge
transfer in chemical reactions, the density $J_{\rm WB}^{}(\omega)$
is not in the simple power-law form (\ref{lowfr}), and also the finite
band width $\omega_c$ may have important influence on the dynamics.
Nevertheless, the associated spectral density $J_{\rm TB}^{}(\omega)$ is
given by (\ref{jjtb}), and  one may now take advantage
of performing numerical simulations of the equivalent TB
model. We wish to emphasize that even the crossover from quantum tunneling to
thermal hopping across the barrier can be studied in the equivalent TB model.

According to the generalization given here, important conclusions
can now be drawn also for non-Ohmic damping. In the sequel, we present
new results obtained by analytic methods for the TB model at low temperatures
and high-frequency cutoff. Upon using (\ref{duality}), we then describe the
implications for the dual WB model.

Consider now first the TB model for super-Ohmic damping. It has been
shown in earlier work \cite{leggett87,weiss91} that the TB model for
$1< s' <2$ is perturbative in $V^2_0$ at finite $T$ and has a
diffusive limit $\mu_{\rm TB}^{}(\omega\to 0) = \bar{\mu}$, where
$\bar{\mu}$ depends on temperature and the leading term is of order $V_0^2$.
With decreasing temperature, the effective transfer matrix element
of the super-Ohmic TB model scales to higher values for fixed
$\omega$, and for $T=0$
and $\omega\to 0$ it even scales to infinity. Hence the model becomes
nonperturbative in $V_0^2$ in this limit. We have been able to perform
the resummation of the full power series in $V_0^2$ \cite{mau}. Interestingly,
we find that the asymptotic behavior is independent of $V_0$ and is
$\mu_{\rm TB}^{}(\omega\to 0) =
(-i\omega/\widetilde{\omega})^{1 - s^\prime} \mu_0^{}$.
Thus, the tight-binding lattice is completely dissolved in this limit, and
the mobility is that of a free super-Ohmic Brownian particle.
In linear response, the mean position grows superdiffusively \cite{gra},
$\langle Q(t)\rangle \propto t^{s^\prime}$.

In contrast, the sub-Ohmic TB model $(s^\prime\!<\!1)$ is nonperturbative in
$V_0^2$ at finite $T$ in the limit $\omega \to 0$. We have been able again
to sum the full power series in $V_0^2$ \cite{weiss91,mau}, yielding
subdiffusive behavior $ \mu_{\rm TB}^{}(\omega\to 0) =
(-i\omega/\widetilde{\omega})^{1-s'} \mu_0^{}$, and the subleading
term is $\propto (1/V_0^2) (-i\omega/\widetilde{\omega})^{2-2s'}$.
The linear mobility vanishes in the dc limit. Regarding the behavior
in the time regime,
in linear response, the mean position $\langle Q(t) \rangle$ grows
sluggishly with subdiffusive power law $t^{s^\prime}$.
As the temperature is decreased, the effective transfer matrix element
of the sub-Ohmic TB model scales to lower values for fixed $\omega$,
and at $T=0$ it vanishes with an essential singularity
$\propto \exp(-{\rm const}/\omega)$ as $\omega\to 0$.
Hence the mobility vanishes faster than any power of $\omega$
in the zero-frequency limit, and the particle gets strictly lo\-calized.

Upon using these results of the TB model and the relation (\ref{duality})
with the mapping $s^\prime \to 2 -s $, we now immediately get new results
for the mobility of the WB model.

The linear mobility of the
sub-Ohmic WB model at finite $T$ is  $\mu_{\rm WB}^{}(\omega\to 0) =
\left(-i\omega/\widetilde{\omega}\right)^{1-s} \mu_0^{}\,$,
and the subleading correction is
$- (-i\omega/\widetilde{\omega})^{2-2s} \bar{\mu}\,$. The
potential represents an irrelevant perturbation in this regime.
The mean position of the particle
$\langle Q(t)\rangle$ grows subdiffusively $\propto t^s$ with $s<1$
\cite{gra}, i.e., just as in the absence of the potential, and the
leading potential corrections grow as $t^{2s-1}$, which is in agreement
with the findings in Ref.\cite{chen}. At $T=0$, the term
$\propto \omega^{1-s}$, representing
the free Brownian particle, is cancelled by the asymptotic expression
resulting from the full power series in $V_0^2$.
As a result, the mean position of the particle grows even slower than
in the free sub-Ohmic case.

In the super-Ohmic WB model at finite $T$, the superdiffusive
contribution of the free Brownian motion is cancelled once again,
and we find slower diffusive behavior,
$\mu_{\rm WB}^{}(\omega\to 0) ={\rm const}$.
On the contrary at $T=0$, the potential is an irrelevant perturbation in
the super-Ohmic WB model. Thus we obtain the superdiffusive characteristics
of the free Brownian particle, $\mu_{\rm WB}^{}(\omega\to 0) =
(-i\omega/\widetilde{\omega})^{1-s}\mu_0^{}\,$ in this limit.

Considering the fact that the system becomes increasingly sensitive to its
low-frequency properties as the temperature is decreased, the above
characteristics are fully consistent with physical intuition.
For sub-Ohmic damping and decreasing temperature, the mobility is
progressively suppressed at constant low frequency in both models.
This is due to the  diverging spectral damping function
$\gamma(\omega)$ for $\omega\to 0$.
On the contrary, for super-Ohmic damping the mobility
is enhanced with decreasing temperature in both models since the
respective spectral damping function vanishes at zero frequency.

The duality relations presented here are directly relevant to
numerical simulation of quantum transport in a continuous periodic
potential for any form of the spectral damping function.
Using the ``bosonized'' language, charge transfer in a Luttinger liquid
through a barrier or impurity \cite{kane} is also described by models of the
form (\ref{hwc}) or (\ref{htb}). Hence, the results
also apply to charge transport in correlated
fermion systems. For instance, the linear conductance $G(\omega)$ of the
impurity scattering problem is
related to the linear mobility $\mu(\omega)$ in the washboard
potential with period $Q_0$ by $G(\omega) = e^2 \mu(\omega)/Q_0^2$.
Further, the equilibrium noise spectrum of the current is given by
$ S(\omega) = \hbar\omega\coth(\hbar\omega/2k_{\rm B}^{}T)G(\omega)\,$.
The interesting case of unscreened Coulomb interactions in the
impurity scattering problem \cite{fabr}, and in tunneling between edge states
of fractional quantum Hall liquids \cite{wen},
corresponds to logarithmically diverging sub-Ohmic behavior,
$\gamma(\omega\to 0) \propto -\ln(\omega/\tilde{\omega})\,$.

In conclusion, we have shown that there exists a general duality
transformation between a quantum Brownian particle in a
continuous washboard potential and a dissipative tight-binding model.
Because the tight-binding approximation ignores excited states at each
lattice site, one might think that
a general mapping between the two models is impossible.
We have shown, however, that the tight-binding
model is fully sensitive to all aspects of both the quantum and the
thermal hopping dynamics of the continuous model. In fact, the tight-binding
model is reconciled with the continuous model by properly taking into
account the change in the spectral density.

We thank R. Egger and L.S. Schulman for valuable discussions.

\end{document}